\documentclass{edp-conf}
\usepackage{graphicx}
\begin{document}

\TitreGlobal{Exact vs. Gauss-Seidel non--LTE radiation transfer solutions}

%%-----------------------------
%%      the top matter
%%-----------------------------
\title{Exact vs. Gauss-Seidel numerical solutions of the non-LTE
radiation transfer problem}
\author{Quang, C.}\address{Observatoire Midi-Pyr\'en\'ees, LA2T, 14
ave. E. Belin, 31400 Toulouse}
\author{Paletou, F.$^1$}
\author{Chevallier, L.}\address{Observatoire de Lyon, CRAL, 9
ave. C. Andr\'e, 69561 Saint-Genis-Laval Cedex}
\runningtitle{Exact vs. Gauss-Seidel non--LTE radiation transfer solutions}
\setcounter{page}{322}
% Keep this line, even if the page will be settled afterwards..
\index{Quang, C.}
\index{Paletou, F.}
\index{Chevallier, L.}
% Repeat the authors here, this will help to make the final index

\maketitle
\begin{abstract}
Although published in 1995, the Gauss-Seidel method for solving the non-LTE radiative
transfer problem has deserved too little attention in the
astrophysical community yet. Further tests of the performances and of
the accuracy of the numerical scheme are provided.
\end{abstract}
%
%%-----------------------------
%%      your text
%%-----------------------------
\section{Introduction}

Fast and {\em accurate} numerical schemes for the solution of the
non--LTE radiation transfer (RT) problem still need to be pushed
further, in order to be able to deal with increasingly complex models
(e.g., multi-dimensional geometry, multi-level atoms,
polarisation...). Hereafter, we present new numerical tests against
{\em exact} solutions of the Gauss-Seidel (GS) and SOR methods
initially proposed by Trujillo Bueno \& Fabiani Bendicho (1995), and
inspired by the classical GS iterative method in numerical analysis
(which can be modified into SOR methods when an overcorrection is made
as compared to GS).

Very few tests, indeed, are available for the validation of any new
numerical scheme dealing with the resolution of the non--LTE radiative
transfer problem. The usual one comes from the computation of
numerical solutions under the Eddington approximation i.e., adopting a
very coarse angular quadrature such as $\mu= \pm 1/ \sqrt{3}$ (also
known as the ``two-stream'' approximation; see \S 4.3.1.  in Rutten
2003). In such a case, analytical solutions -- of a ``reduced'' RT
problem though -- are available for numerical solutions to be checked
against. However this test may lead to an {\em erroneous} estimation
of the accuracy of the numerical scheme, as pointed out in Chevallier
et al. (2003, see \S 5. therein).

\section{Numerical tests and discussion}

\begin{figure}
\centerline{
\begin{tabular}{cc}
\includegraphics[width=6.3cm]{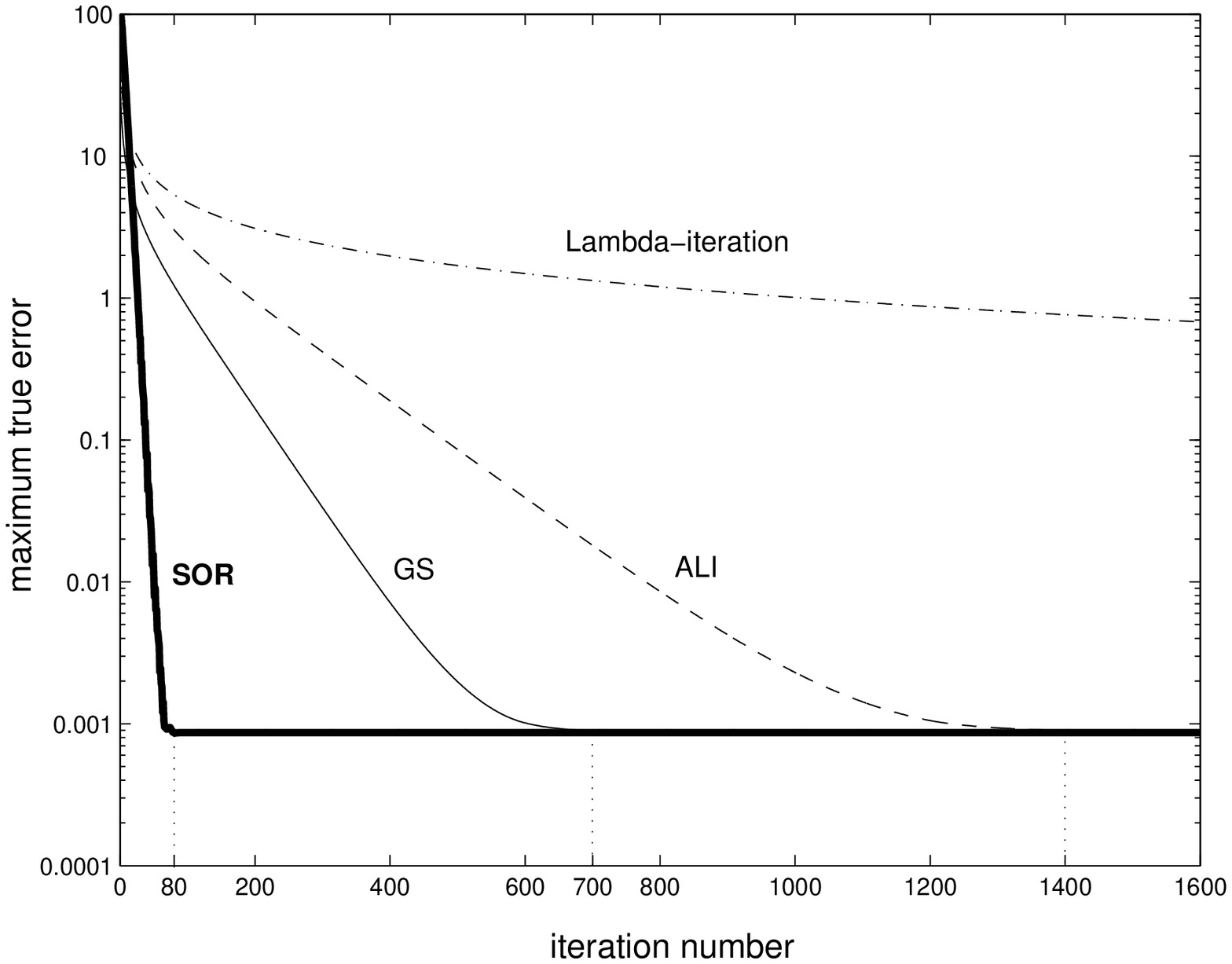} & \includegraphics[width=6.3cm]{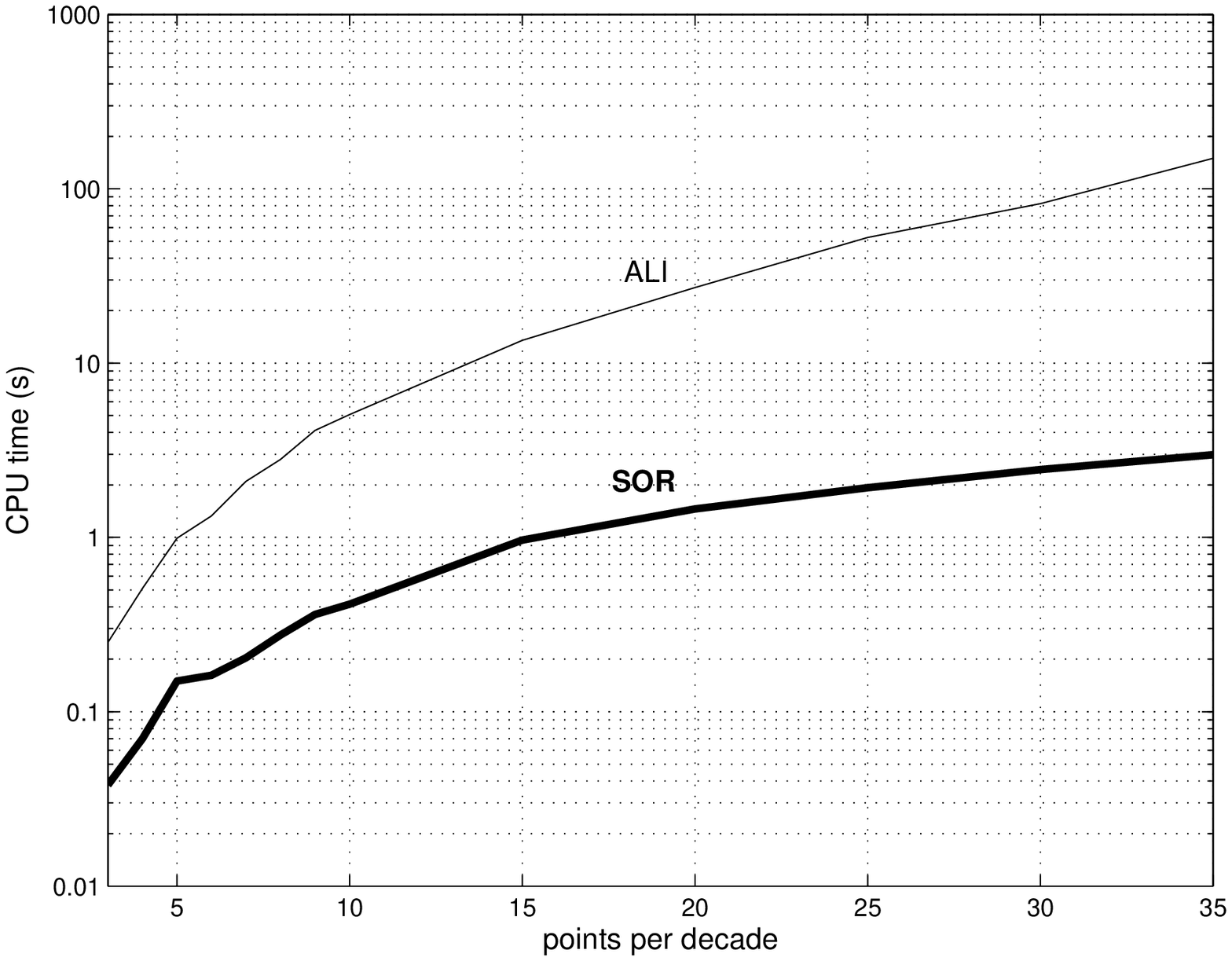}
\end{tabular}}

\caption{{\em Left:} evolution of the maximum true error vs. the
number of iterations (20 grid-points per decade) {\em Right:} CPU-time
for, respectively, the ALI and the GS/SOR numerical schemes vs. the
number of points per decade (for a 1D finite slab of optical thickness
1000 and a collisional destruction probability $10^{-4}$).}

\label{example}
\end{figure}

The GS/SOR numerical schemes are better implemented when one adopts
the ``short characteristics'' (SC) approach for the so-called formal
solution of the RT equation (see Auer \& Paletou 1994). Chevallier et
al. (2003) showed how the accuracy of an Accelerated
$\Lambda$--Iteration (with a monotonic parabolic SC formal solver)
numerical code's solutions do scale with the refinement of both
spatial and angular grids, by comparing the latter to very
high-accuracy analytical solutions given by the ARTY code (Chevallier
\& Rutily 2004).

We computed GS/SOR solutions for a grid of 1D finite slab, two-level
atom (in CRD) models that we compared to ARTY reference solutions: the
rate of convergence for various numerical schemes is displayed in
Fig.~1 (left). Our main conclusions are that: (a) as expected, the
{\em accuracy} of the GS/SOR code is identical to the one of the
ALI-SC-based code, (b) the numerical ``overcost'' for GS iterations
(due to a modified formal solver) is {\em negligible} but, (c) that
the gain in computing time with GS/SOR is {\em very significant.} As
shown in Figs.~1, for high-order quadratures, which are {\em absolutely
needed} in order to keep the accuracy of the ALI-SC method better than
1\% (see Chevallier et al. 2003), the gain in CPU-time provided by the
GS/SOR scheme can be as large as a factor of 50! And even with standard
acceleration techniques for ALI or GS, GS/SOR remains the fastest
algorithm (see Trujillo Bueno \& Fabiani Bendicho 1995).

We feel that this fully justifies to consider {\em very seriously}
GS/SOR schemes for future radiative modelling codes (note also that
GS/SOR have already been generalized to complex models). It is
particularly important for the development of those diagnosis tools
required by major projects such as GAIA or HERSCHEL, for instance.

%%-----------------------------
%%      your bibliography
%%-----------------------------

\end{document}